\documentclass[twocolumn,journal]{IEEEtran}
\usepackage{cite}
\usepackage{amsmath,amssymb,amsfonts}
\usepackage{algorithmic}
\usepackage{graphicx}
\usepackage{adjustbox}
\usepackage{textcomp}

 \usepackage{xcolor}   
 \newcommand{\rev}[1]{{\color{black} #1}} 
 \newcommand{\black}[1]{{\color{black} #1}}

\begin{document}

\title{Significant Low-dimensional Spectral-temporal Features for Seizure Detection}

\author{Xucun~Yan,
        Dongping Yang,
        Zihuai Lin,~\IEEEmembership{Senior Member, IEEE,}
        and~Branka Vucetic,~\IEEEmembership{Fellow,~IEEE}
\thanks{X. Yan, Z. Lin and B. Vucetic are with School of Electrical and Information Engineering, University of Sydney, New South Wales 2006, Australia (e-mail: xucun.yan@sydney.edu.au; zihuai.lin@sydney.edu.au;branka.vucetic@sydney.edu.au).}
\thanks{D. Yang is with Quanzhou Institute of Equipment Manufacturing, Haixi Institute, Chinese Academy of Sciences, Quanzhou 362200, and Fujian Institute of Research on the Structure of Matter, Chinese Academy of Sciences, Fuzhou 350002, China(e-mail: dpyang@fjirsm.ac.cn).}
}

\maketitle
\begin{abstract}
Absence seizure as a generalized onset seizure, simultaneously spreading seizure to both sides of the brain, involves around ten-second sudden lapses of consciousness. It common occurs in children than adults, which affects living quality even threats lives. Absence seizure can be confused with inattentive attention-deficit hyperactivity disorder since both have similar symptoms, such as inattention and daze. Therefore, it is necessary to detect absence seizure onset. However, seizure onset detection in electroencephalography (EEG) signals is a challenging task due to the non-stereotyped seizure activities as well as their stochastic and non-stationary characteristics in nature. 
Joint spectral-temporal features are believed to contain sufficient and powerful feature information for absence seizure detection.
However, the resulting high-dimensional features involve redundant information and require heavy computational load.
Here, we discover significant low-dimensional spectral-temporal features in terms of mean-standard deviation of wavelet transform coefficient (MS-WTC), based on which a novel absence seizure detection framework is developed.
The EEG signals are transformed into the spectral-temporal domain, with their low-dimensional features fed into a convolutional neural network. 
Superior detection performance is achieved on the widely-used benchmark dataset as well as a clinical dataset from the Chinese 301 Hospital.
For the former, seven classification tasks were evaluated with the accuracy from 99.8\% to 100.0\%, while for the latter, the method achieved a mean accuracy of 94.7\%, overwhelming other methods with low-dimensional temporal and spectral features.
Experimental results on two seizure datasets demonstrate reliability, efficiency and stability of our proposed MS-WTC method, validating the significance of the extracted low-dimensional spectral-temporal features. 
\end{abstract}


\section{Introduction}
\label{introduction}
Epilepsy, as one of the most common neurological diseases, affects around 50 million population in the world~\cite{geng2020epileptic}. It is characterized by excessive and sudden electrical discharge, which is generally identified by manually scanning long-time EEG signals with commonly time consuming, error prone and with low consistency between physicians. Thus, it is important to develop a reliable, effective and stable automatic seizure detection system based on EEG signals, paving the way for technologies of online seizure onset detection as well as further intervention. However, it is a challenging task due to the stochastic and non-stationary properties of EEG signals in nature, although a lot of efforts have been spent in this field \cite{d8,d14,d20,zhu2021cognitive}.

A number of seizure detection methods have been proposed based on the data features in the temporal, spectral, and spectral-temporal domain \cite{8911225,d76,d70}.
Meanwhile, various datasets are created and utilized to develop and examine these methods \cite{d66,d67,d23}.

Multiple methods have been proposed to extract temporal features~\cite{d8,d1,d2,d3}. 
Lin et al. \cite{d1} designed a headband for epileptic seizure detection, accompanied by the approximate entropy (ApEn) for ictal and inter-ictal classification; empirical mode decomposition (EMD) has been employed to decompose EEG signals into intrinsic mode functions, which are fed into a least square support vector machine (LS-SVM)~\cite{d2} or an artificial neural network (ANN)~\cite{d3} for seizure detection.
However, these features have been demonstrated not enough to provide an effective way for recognizing the intrinsically rhythmic epileptic EEG signals \cite{d73}, leading to a poor recognition performance.

Various rhythmic characteristics of epileptic EEG signals have also been investigated~\cite{d76,d77}. 
Fast Fourier transform (FFT) is used to compute Welch power spectral density (PSD), which is fed into a decision tree classifier~\cite{d76} or an artificial immune recognition system with fuzzy resource allocation mechanism classifier~\cite{d77}.
However, these studies used restricted number of features in a single-domain and thus only achieved limited success.
Meanwhile, it is generally perceived that Fourier methods are not suitable for the non-stationary signal processing~\cite{d74}. 

To better capture significant features from non-stationary EEG signals, joint spectral-temporal feature learning framework has been proposed~\cite{d70,d19,d12,d13,d10,d11,d75}.
Both continuous wavelet transform (CWT) and discrete wavelet transform (DWT) are employed to extract the spectral-temporal features and obtain a time-frequency image~\cite{d70,d19}.
The enhanced method of DWT--dual-tree complex wavelet transform \cite{d12,d13} has also been used to classify healthy, epileptic seizure-free and epileptic seizure signals~\cite{d14,d4,d9,d15}. 
The short time Fourier transform (STFT) was utilized to get multi-view spectrum~\cite{d10} and PSD~\cite{d11} for seizure detection.
Significantly, a recent work by Nhan et al. \cite{d75} applied STFT to generate a similar image for seizure prediction.
The aforementioned works indicate that such kind of time-frequency images contains sufficient and powerful feature information.
However, their high dimension features are too complicated and expected to involve redundant information, leading to large computation time and potential overfitting.
Thus, finding significant spectral–temporal features for seizure detection remains an open problem.

The other issue is the dataset for validation. 
Most schemes utilized the published datasets \cite{d66,d67,d23}, which are collected for research purpose rather than clinical use. 
The former is constrained by specific conditions, such as equipment, subjects and various collection requirements. 
Existing approaches~\cite{d49,d50,d51,d52} have already achieved good results (over 99\% accuracy) on such kind of data, e.g., the published data from Bonn University \cite{d23}.
However, the raw clinical EEG data involving artifacts may reduce the performance. 
Artifacts are inevitable in the clinical EEG data, which is interfered by the electromyogram (EMG) and electrocardiogram (ECG) signals. 
Generally, existing methods based on published datasets cannot be popularized in practice~\cite{louis2016normal}. 

To address these issues, we calculate mean-standard deviation of wavelet transform coefficients (MS-WTC) obtained by a CWT method~\cite{d68,d72}. The method is capable to reduce the feature dimension and attenuate the artifact interruption~\cite{louis2016normal} for the non-stationary EEG signal analysis.
This method is employed to effectively detect the clinical seizure EEG signals with the following main procedure. 
Firstly, EEG signals are segmented and filtered to 4s samples.
Secondly, each sample signal is transformed by using CWT to get the time-frequency image, which will be expressed as a matrix quantifying PSD series for various time points and time scales. 
Then, the time-averaged PSD and its standard deviation is
calculated to get the low-dimension features at each time scale.
Finally, such features are split into training and testing sets and
fed into a convolutional neural network (CNN) for seizure detection.
By comparing to other methods with low-dimensional temporal and
spectral features, a competitive seizure detection method on both
Bonn database and a dataset from the Chinese 301 Hospital
is achieved.

\section{Databases}

\label{Datasets}
\subsection{Description of \rev{Bonn dataset}\label{sec:data_A}}

\begin{figure}[htbp]
  \centering
  \includegraphics[width=\columnwidth,trim={0cm 0.815cm 0cm 0cm},clip]{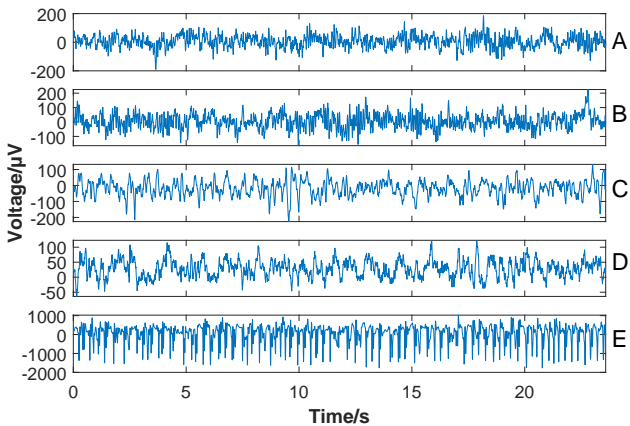}
  \caption{EEG signals of set A, B, C, D and E from \rev{Bonn dataset}}
  \label{fig:f1}
\end{figure}

\begin{table}[htbp]
\caption{Cases considered for classification on \rev{Bonn dataset}}
\begin{adjustbox}{width=\columnwidth,center}
\centering
\begin{tabular}{cccccccc}
\hline
\hline	
Classes& Case 1& Case 2& Case 3& Case 4& Case 5& Case 6& Case 7  \\
\hline
Seizure& E& E& E& E& E& E& E  \\
Non-seizure& A& B& C& D& ACD& BCD& ABCD  \\
\hline
\hline	
\end{tabular}
\label{tab_1}
\end{adjustbox}
\end{table}

The database consisting of five subsets (A - E) is available in \cite{d23} (see Fig. \ref{fig:f1}). 
These subsets are collected from a 128-channel amplifier with a sampling frequency of 173.61 Hz \cite{d23}. 
\rev{Intracranial electrodes are attached}, and subjects are in an awake state, applied with 10-20 international standard systems \cite{d65}. 
Each subset consists of 100 samples with a duration of 23.6 s for each sample. There are 4096 points in each sample and the bandwidth is 86.8 Hz \cite{d58}. 
Usually, the data can be denoised through a bandpass filter with cut-off frequencies of 0.53 Hz and 40 Hz \cite{d23}. 

Subsets A and B are taken from healthy volunteers with eyes open and closed, respectively. 
Subsets C, D and E are recorded from epileptic patients. 
Subset C is collected by recording the signals in hippocampal formation in the opposite hemispheres to the epileptogenic zones while subset D is from the epileptogenic zones \cite{d65}. 
Both subsets C and D measure patients under a seizure-free condition. 
Data in subset E is acquired in lateral and basal regions of the neocortex when seizure occurs \cite{d23}. 
Seven common cases are explored in this paper with details listed in Tab. \ref{tab_1}. 

\subsection{Description of \rev{C301 dataset}\label{data:B}}

\begin{table}[htbp]
\caption{Data information of \rev{C301 dataset}. 
$T_{min}$ and $T_{max}$ in the second
column denote the minimum and maximum seizure durations of seizure events for each subject.}
\begin{adjustbox}{width=\columnwidth,center}
\centering
\begin{tabular}{cccc}
\hline
\hline	
ID	& Seizure events ($T_{min}$ - $T_{max}$) (s)	& Total Seizure time (s)	& Total seizure-free time (min)	\\
\hline

1   &   26  (7.4 - 84.4 ) 	& 980.0 	& 34.5 	\\
2	&   53  (2.0 - 21.8 ) 	& 300.8 	& 45.8 	\\
3	&   20  (12.1 - 45.3 ) 	& 499.8 	& 42.5 	\\
4	&   16  (1.6 - 45.3 ) 	& 89.4   	& 49.3 	\\
5	&   10  (1.6 - 8.3 ) 	& 45.3 	    & 50.1 	\\
6	&   6   (5.1 - 14.1 ) 	& 55.5   	& 49.9 	\\
7	&   1   (7.8 - 7.8 ) 	& 7.8    	& 50.7 	\\
8	&   12  (3.5 - 8.6 ) 	& 78.3   	& 49.5 	\\
9	&   5   (3.1 - 18.6 ) 	& 51.4   	& 50.0 	\\
10	&   17  (1.2 - 12.5 ) 	& 129.2  	& 48.7 	\\
11	&   10  (2.3 - 11.3 ) 	& 80.8   	& 49.5 	\\
12	&   8   (2.0 - 7.2 ) 	& 44.3   	& 50.1 	\\
13	&   6   (2.7 - 9.4 ) 	& 34.4   	& 50.2 	\\
14	&   16  (2.3 - 42.1 ) 	& 179.8 	& 47.8 	\\
15	&   20  (2.3 - 13.7 ) 	& 133.5 	& 48.6 	\\
16	&   20  (2.3 - 11.7 ) 	& 133.5 	& 48.6 	\\
17	&   12  (2.0 - 15.6 ) 	& 95.4   	& 49.2 	\\
18  &   8   (1.6 - 7.4 ) 	& 41.1   	& 50.1 	\\
19	&   12  (8.6 - 22.5 ) 	& 206.4  	& 47.4 	\\
\hline
Sum &   258           & 3186.6  &912.4 \\

\hline	
\hline
\end{tabular}
\label{Data information of dataset B}
\end{adjustbox}
\end{table}

\begin{figure*}[h]
  \centering
  \includegraphics[width=\textwidth,trim={0cm 0.9cm 0 0.1cm}, clip]{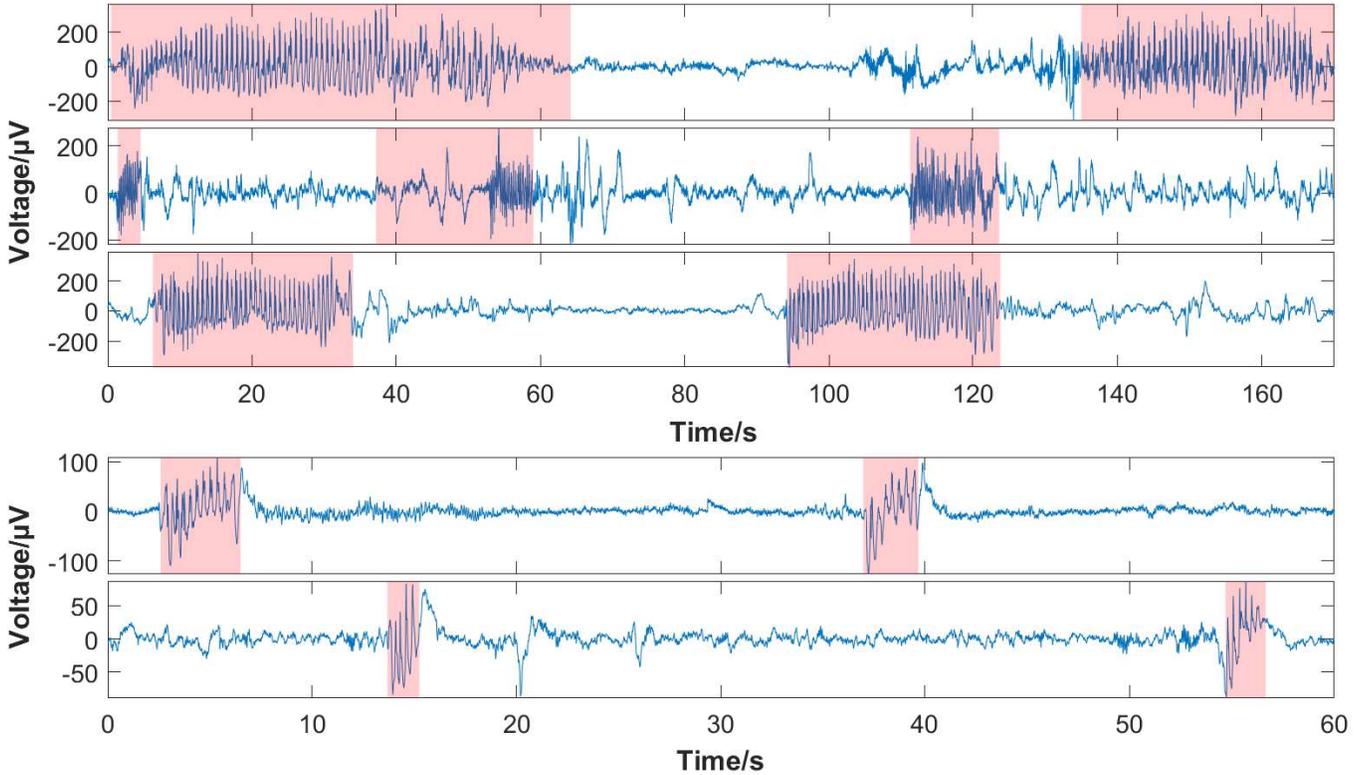}  
  \caption{EEG signals of five subjects from \rev{C301 dataset}. Red shadows indicate seizure regions marked by physicians.}
  \label{fig:2}
\end{figure*}

This dataset is collected from the Chinese 301 Hospital, \rev{in short C301 dataset}, with 19 absence seizure subjects. 
Each patient's EEG signals are recorded with 19-channel electrodes attached to the scalp.
The sampling frequency $f_s$ is 256 Hz.
More information on \rev{C301 dataset} is shown in Tab. \ref{Data information of dataset B}, where one can find that the seizure event numbers vary from 1 to 53 over different subjects and the seizure durations are widely distributed.
For each subject, the seizure-free time is generally dozens (around 17 times in average) of the seizure time, leading to a larger data size unbalance for different classes than the one of \rev{Bonn dataset}.

Moreover, the clinical EEG signals of both seizure and non-seizure are diverse and complicated, as shown in Fig.~\ref{fig:2} with seizure regions red shadowed by Physicians.
From subject to subject, the EEG signals for seizures vary in terms of durations as well as wave patterns, while non-seizure patterns are also diverse.
From the clinical EEG signals, one can find the non-stereotyped seizure activities as well as their stochastic and non-stationary characteristics, leading to the difficulties in seizure detection.

\section{Methodology\label{Methodology}}
\subsection{\label{sec:data_pre}Data preparation}

\begin{figure}[htbp]
  \centering
  \includegraphics[width=\columnwidth]{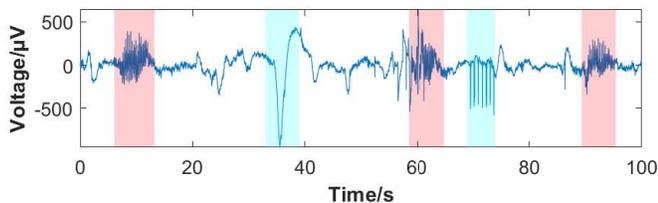}  
  \caption{EEG signals from \rev{C301 dataset} with seizures and artifacts.
  Seizure regions are red shadowed and artifacts are cyan shadowed.}
  \label{f2}
\end{figure}

The typical absence seizure signals consist of periodic spike-waves, which can be easily captured at a specific frequency band. 
However, the clinical EEG signals are often interfered by physiological (e.g., induced from ECG, pacemaker, eye movements and sniffing) or non-physiological (e.g., from bed or chair movements and dropped electrodes) artifacts \cite{d21,d24}.
In \rev{C301 dataset}, one can find such kind of artifacts in Fig.~\ref{f2}, which will be addressed with our MS-WTC method to be introduced in Sec.~\ref{sec:features}.

There are 19-channel electrodes in \rev{C301 dataset}, only one of which is selected due to the high spatial coherence in the absence seizure EEG signals.
For \rev{Bonn dataset}, the samples with a duration of 23.6s for each one are mentioned in Sec.~\ref{sec:data_A}, while for \rev{C301 dataset}, the continuous EEG recordings are segmented into half-overlapping 4s samples, resulting in $28964$ samples in total.
Each sample is labeled as $y_i$ for the $i$-th sample, with $y_i=1$ for the seizure-dominant sample and $y_i=0$ for the non-seizure-dominant sample.

\subsection{\label{sec:features}Data processing}

\begin{figure}[h]
  \centering
  \includegraphics[width=\columnwidth,trim={6.5cm 0.5cm 7.8cm 0cm},clip]{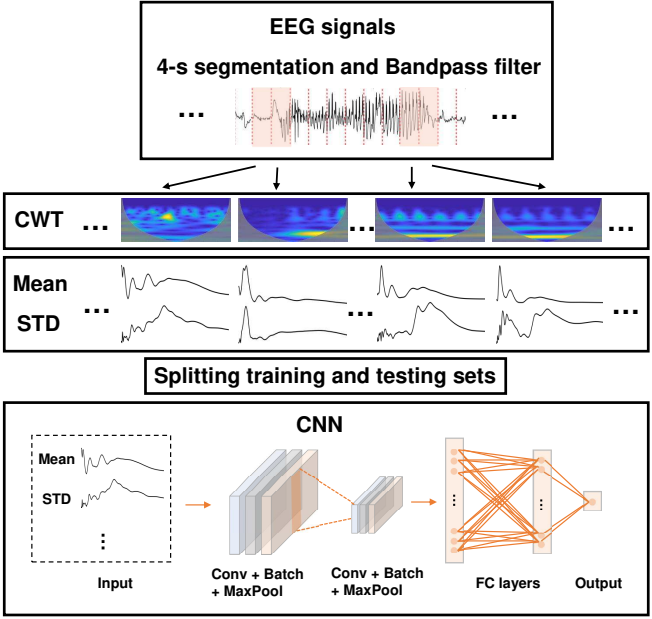}  
\caption{The flowchart of the proposed seizure detection framework}
\label{process}
\end{figure}

Figure~\ref{process} depicts the procedure of seizure detection framework, including data preprocessing, our MS-WTC method and neural network classification. 
In the preprocessing, each data sample is filtered with the commonly used frequency band (0.1-35 Hz), and the data section in some samples with abnormal large amplitudes (exceeding 3 times of standard deviation of the sample) are set to the baseline (voltage $= 0$). 
Then one can get the preprocessed sample $\boldsymbol{x_i}$.
In the proposed MS-WTC method, $\boldsymbol{x_i}$ is transformed into a time-frequency image through CWT~\cite{d30}, where the morse wavelet with the default setting is used as the mother wavelet~\cite{d33}. 
\rev{Generalized Morse wavelet family is useful for analyzing modulated signals and localized discontinuities \cite{olhede2002generalized,lilly2012generalized} and is therefore suitable for detecting the stereotyped spike-wave EEG signals of absence seizure activities. Other typical wavelets, such as Daubechies, Bump, Morlet, etc., are also tested with only slight changes of detection capability.}

As commonly performed in the aforementioned works~\cite{d19,d70,d10, d11,d75}, the time-frequency image can be expressed as a matrix $\boldsymbol{M_i}$ quantifying PSD series for various time points and time scales as follows
\begin{equation}
    \boldsymbol{M_i}=
  \left[ {\begin{array}{ccccc}
  m_{1,1} & m_{1,2} & ... & m_{1,b}& ...\\
  m_{2,1}& m_{2,2} & ... & m_{2,b} & ...\\
   ... & ... & ... & ... & ...\\
   ... & ... & ... & m_{a,b} & ...\\
   ... & ... & ... & ... & ...\\
  \end{array} } \right], 
  \text{ for } 
  \begin{cases}
   1\leq a \leq k,\\
   1\leq b \leq l,
  \end{cases}
\end{equation}
where $m_{a,b}$ are the transformed coefficients, $k$ is the number of scales and $l=4f_s$ is the sample size.
By definition, $k$ depends on the $f_s$, the lowest and highest peak band-pass frequencies (0.1-35 Hz) \cite{d68}.  
In the matrix, the coefficients in $a$-th row represent the time series resulting from the convolution of the signal $\boldsymbol{x_i}$ with the morse wavelet at the $a$-th time scale (or frequency).

As discussed in the introduction, our aim is to effectively extract the significant low-dimensional spectral-temporal features from non-stationary EEG signals.
To this end, we here simply employ the time-averaged PSD magnitude $\mu_a$ as well as its standard deviation $\sigma_a$ at $a$-th time scale, given as
\begin{align}
\label{eqn:a}
\mu_a &= \frac{1}{l_a}\sum_{b} m_{ab},\\
\label{eqn:b}
\sigma_a &= \sqrt{\frac{1}{l_a}\sum_{b} (m_{ab}-\mu_a)^2
},
\end{align}
where $a$ ranges from 1 to $k$ and the range of $b$ is confined by the cone of influence to exclude the boundary effects with the total number of coefficients at $a$-th row given as $l_a$~\cite{d79}.

One may suspect that only one of these two features can achieve a considerable performance.
To this end, we also investigate other two methods using either one of them.
One is the mean of wavelet transform coefficient (M-WTC) method, and only $\mu_a$ is used. The other is the standard deviation of wavelet transform coefficient (S-WTC) method, and only the $\sigma_a$ is applied.

To demonstrate the significance of these extracted features, we compare our method with the ones using other low-dimensional features, such as the temporal features by EMD and the spectral feature by FFT. 
The high-dimensional spectral-temporal features by CWT is not employed here due to large computational requirements and its cone-shape image with boundary effects excluded, which will be our future work.

\subsection{Training and classification}
\label{Training and classification}
We here employ the CNN model for the training and classification.
The above features are input into the first convolutional layer with the channel number adapted to feature dimension. For the methods M-WTC, S-WTC and FFT, the input channel numbers are all set to 1, while for the methods MS-WTC and EMD, they are set to 2 and 4, respectively.
The method EMD decomposes the EEG signals into IMFs as the input features.

The CNN structure is shown in the bottom figure of Fig.~\ref{process} with two convolutional layers, each of which is followed by a batch normalization and a max-pooling. 
The dropout is introduced to reduce the overfitting issue \cite{d42}. 
The kernel size is $3$ and the defaulted stride size is $1$, while the output channel number of the two convolutional layers are set to 32 and 16, respectively.
Then two fully connected layers are added with 
\rev{activation function} Relu and Sigmoid, respectively.
The model details are summarized in Tab.~\ref{CNN summary} by taking the outputs of preprocessing method MS-WTC as input features of the CNN model, where the input shape is specified to (163, 2); Conv1D stands for one dimensional convolutional layer; Max pooling1D stands for one dimensional max pooling. \rev{In fact, before we decide to use 2-layer CNN, SVM and ANN are also tested. In comparison, CNN model performs the best. Additionally, because of limited dimensions of the input features, a deeper CNN model will cause overfitting issue. The good performance discussed in \ref{Results for database B} and \ref{RA} also demonstrates that 2-layer CNN is a proper model structure for accurate seizure detection.}

\begin{table}[htbp]
\caption{Summary of the CNN model}
\begin{adjustbox}{width=3 in,center}
\centering
\begin{tabular}{cc}
\hline
\hline	
Layer name 	                & Output Shape from the layer 	    \\
\hline

Conv1D           &(None, 161, 64)         \\     

Batch normalization  &(None, 161, 64)     \\    

Max pooling1D   &(None, 80, 64)           \\      

Conv1D           &(None, 78, 32)           \\    

Batch normalization   &(None, 78, 32)         \\      

Dropout           &(None, 78, 32)             \\       

Max pooling1D   &(None, 39, 32)             \\      

Flatten           &(None, 1248)               \\      

Dropout            &(None, 1248)                \\     

Dense                 &(None, 10)                   \\ 

Dense               &(None, 1)                     \\

\hline	
\hline
\end{tabular}
\label{CNN summary}
\end{adjustbox}
\end{table}

The loss function is defined as a binary cross-entropy \cite{d43}
\begin{equation}
\label{eqn:e}
L = -\frac{1}{S_o}\sum_{i = 1}^{S_o} [y_i \times \log(\hat{y_i})+ (1-y_i)\times \log(1-\hat{y_i})],
\end{equation}
where $S_o=2$ denotes the number of classes (seizure and non-seizure) and $\hat{y_i}$ is a predicted label for the $i$-th sample. 
Adam optimizer is employed to optimize $L$.
Each database is randomly split into 70\% of samples for training and 30\% for testing, and 100-cross validation is applied to get a relatively stable and robust result. An artificially balanced dataset is generated by randomly selecting equal numbers of samples for each cross validation to avoid trivial solutions, due to the large unbalanced data sizes for different classes in \rev{C301 dataset} as mentioned in Sec.~\ref{data:B}. 
\rev{
It should be noticed that subjects in the training set are not distinct from the subjects of the tests set for \rev{C301 dataset}. Therefore, we actually tested two scenarios, named as mixed scenario and separated scenario to complete research results. The mixed scenario is described as aforementioned, and the separated scenario is that samples are picked from random 6 subjects as testing set and 13 subjects as training set. }

These works are implemented through MATLAB version R2019a for feature extraction and Python 3.7 for model training. 
The computation is performed with 2.2 GHz, 64 GB of RAM and Intel(R) Core (TM) i7-8750H. 
The computation time for predicting a 4s sample roughly spends 1.27 ms $\sim$ 1.48 ms using our proposed method.

\subsection{Measurements}
The binary classification performance of various methods can be evaluated by standard statistical measurements: specificity (SPE), sensitivity (SEN) and accuracy (ACC), defined as
\begin{align}
\text{SPE} &= \frac{\text{TN}}{\text{TN}+\text{FP}}, \\
\text{SEN} &= \frac{\text{TP}}{\text{TP}+\text{FN}}, \\
\text{ACC} &= \frac{\text{TP}+\text{TN}}{\text{TP}+\text{FP}+\text{TN}+\text{FN}}, 
\label{eq:8}
\end{align}
where TP (true positive) is the number of samples correctly predicted as seizure class, TN (true negative) means the number of correctly predicted as non-seizure class, FP (false positive) indicates the number of samples incorrectly detected as seizure class, and FN (false negative) denotes the number of samples incorrectly detected as non-seizure class.
ACC is the general measurement of the correctly predicted ratio of the total testing samples for each dataset.

\section{Experimental results and discussions}
\label{Results}

In Sec.~\ref{sec:Extracted features}, significant low-dimensional spectral-temporal features are presented, while other temporal and spectral features are also displayed for comparison. 
The significance of these extracted features are examined on both databases in Sec.~\ref{RA} and Sec.~\ref{Results for database B}, respectively.
The advantages of our proposed methods are also discussed in each subsection.

\subsection{Low-dimensional spectral-temporal features}
\label{sec:Extracted features}

\begin{figure}[tp]
  \centering
  \includegraphics[width=0.5\textwidth,trim={0 0 0 0}, clip]{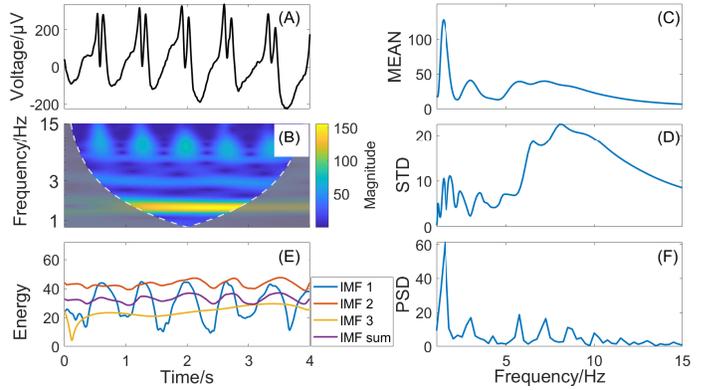}  
  \caption{One example of seizure EEG signals from \rev{C301 dataset} and its temporal, spectral and spectral-temporal features. 
  (A) An original 4s seizure sample; 
  (B) The cone-shaped time-frequency image produced by CWT;
  (C) The time-averaged PSD magnitude ($\mu_a$);
  (D) The PSD standard derivation ($\sigma_a$);
  (E) The energies of the first IMFs and their weighted sum produced by EMD to decompose the EEG signal in (A);
  (F) The PSD produced by FFT.
  }
  \label{f3}
\end{figure}

\begin{figure}[tp]
  \centering
  \includegraphics[width=0.5\textwidth,trim={0 0 0 0}, clip]{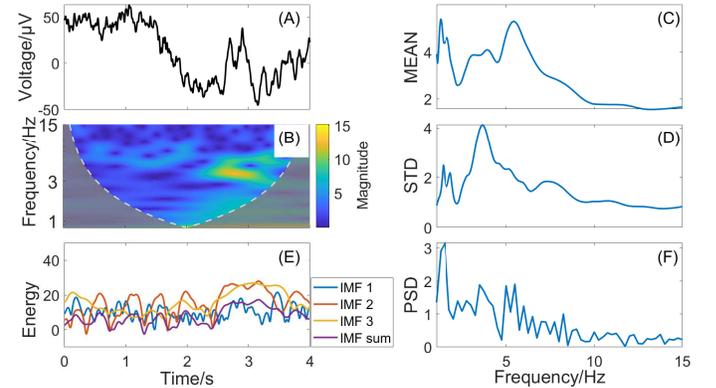}  
  \caption{One example of non-seizure EEG signals from \rev{C301 dataset} and its temporal, spectral and spectral-temporal features. 
  The description of each subplot is the same as that in Fig.~\ref{f3}.}
  \label{f4}
\end{figure}

Here we take the samples of seizure and non-seizure signals from \rev{C301 dataset} as our examples for the feature analysis.
The ones from \rev{Bonn dataset}, which is exhaustively studied, will not be shown. 
The features extracted via various methods are presented in Figs.~\ref{f3} and \ref{f4}.

For the seizure sample in Fig.~\ref{f3}(A), the EEG signal exhibits periodic 2-spike-waves with $\sim1.3$ Hz recurrences.  
The signal is transformed using CWT into the time-frequency image with boundary regions excluded, leading to a cone-shape 2D image, as shown in Fig.~\ref{f3}(B).
The image clearly shows the highest PSD at the frequency $\sim1.3$ Hz for the whole data segment.
Due to the regularity of the spike-wave pattern, PSD peaks also show up at the harmonic frequency regions.
Significantly, the spikes in the spike-wave pattern can further induce the periodic PSD in the higher order harmonic frequency region.
To this end, the main features in the time-frequency image can be summarized into the following two low-dimensional features: the time-averaged PSD magnitude (MEAN) and its standard deviation (STD), which are shown in Figs.~\ref{f3}(C-D).

Besides, the EEG signal's temporal and spectral features are also presented in Fig.~\ref{f3} (E-F) obtained by EMD and FFT, respectively.
EMD has an ability to adaptively decompose the non-stationary signals, capturing the envelope of the signals, presented as the so-called IMF.
Here we employ the first 3 IMFs and their weighted sum as temporal features, where the energy of the first IMF displays a conspicuous oscillation (Fig.~\ref{f3}(E)).
The spectrum by FFT also shows a sharp peak at the frequency $\sim1.3$ Hz, while its harmonic spectral pattern is irregular (Fig.~\ref{f3}(F)). 

The non-seizure EEG signal fluctuates stochastically and non-stationarily without clear periodicity or a characterized pattern (Fig.~\ref{f4}(A)).
The corresponding time-frequency image is irregular with much lower PSD (Fig.~\ref{f4} (B)).
Both of them are contrastive with those of the seizure sample (Fig.~\ref{f3}(A-B)).
As a result, its low-dimensional spectral-temporal features are distinct in seizure (Fig.~\ref{f3}(C-D)) and non-seizure samples (Fig.~\ref{f4}(C-D)).
The properties of the first 3 IMFs are much more complicated (Fig.~\ref{f4}(E)), compared to that of the seizure sample (Fig.~\ref{f3}(E)).
For the spectral feature by FFT, there is also a weak peak at the low frequency and similar PSD distribution (Fig.~\ref{f4}(F)) with that of seizure sample (Fig.~\ref{f3}(F)), which is in contrast with the distinction of the time-averaged PSD distribution as well as its standard deviation distribution by CWT for seizure (Fig.~\ref{f3}(C-D)) and non-seizure samples (Fig.~\ref{f4}(C-D)).
These indicate that our proposed low-dimensional spectral-temporal features should improve the performance of seizure detection, to be demonstrated on both databases in the following subsections.

Moreover, our MS-WTC method takes the statistical information of time-frequency image at each scale, preserving the ability to capture the intrinsic properties of non-stationary signals and effectively reducing the feature dimension by removing the redundant information involved in the 2D time-frequency image produced by CWT. 
\rev{As well known, the time series signal can be expanded into and represented by various moments. Specifically, the first and second order moments contain 100\% information of Gaussian signals. Surprisingly, it has been discovered that the first and second order moments of neuronal spiking signals dominate more than 90\% information of the real retinal activities \cite{schneidman2006weak}. Therefore, the first and second order moments-mean and variance (standard deviation in our study), as the most significant features, are expected to take sufficient information of the original signals for further processing.} This compact statistical information reduces the computation complexity and heavy load to significantly facilitate big data analysis. \rev{The sensitivity analysis is done by comparing different results from the first order moment (M-WTC), the second order moment (S-WTC) and combined first and second order moments (MS-WTC). The results shown in the Section \ref{RA} and \ref{Results for database B} indicate that the combined first and second order moments, MS-WTC, has the best performance.}

\subsection{Results for \rev{Bonn dataset}}
\label{RA}

\begin{figure*}[h]
  \centering
  \includegraphics[width=\textwidth,trim={0.5cm 0.75cm 0.2cm 0.9cm}, clip]{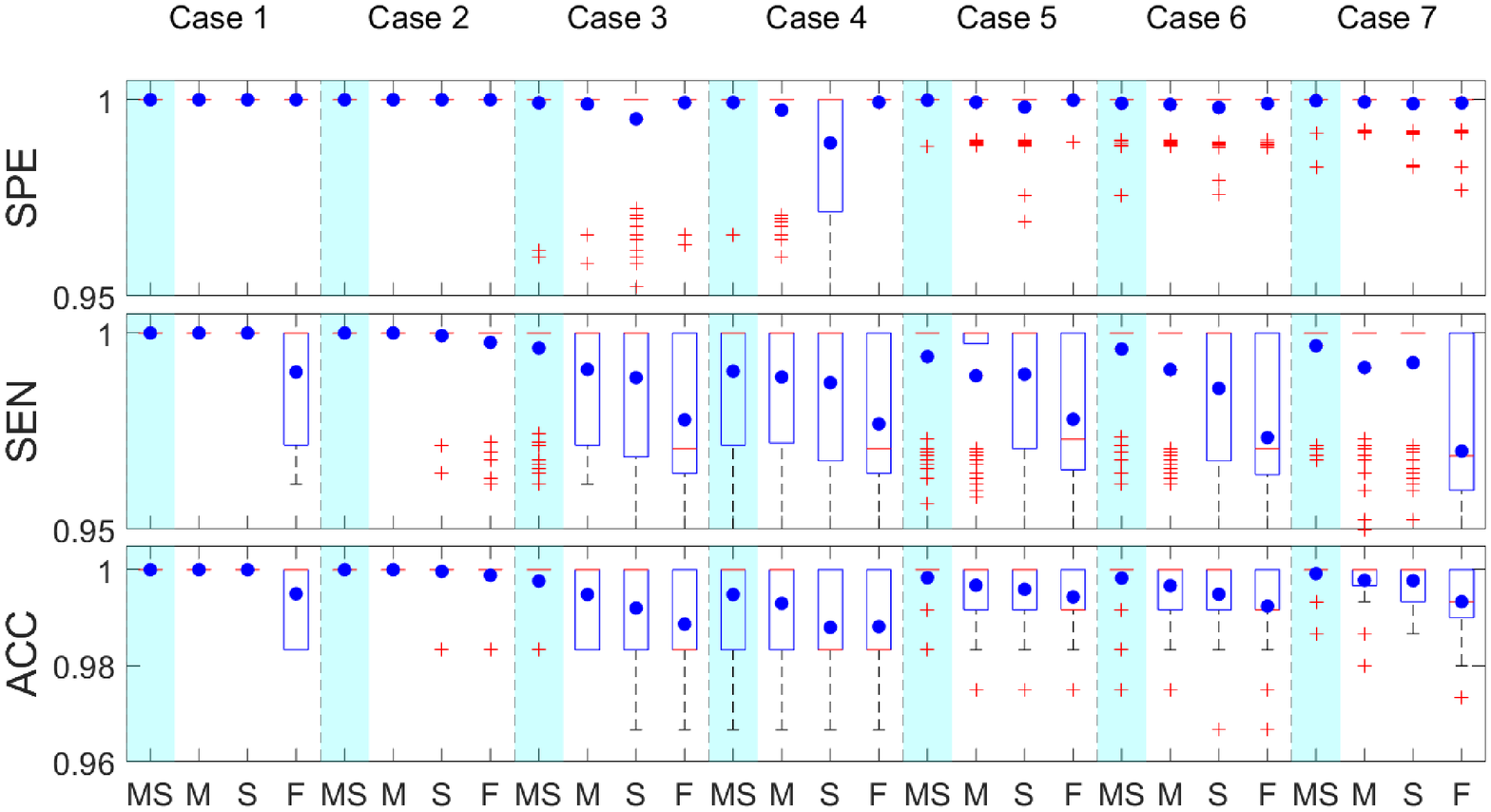} 
  \caption{Evaluations of various methods with 100-cross validation on the seven classification cases of \rev{Bonn dataset}.
  The standard statistical measurements: SPE, SEN and ACC are used for evaluation of the methods: MS-WTC (MS), M-WTC (M), S-WTC (S) and FFT (F);
  Box plot is employed with the box top and bottom denoting the the $75_{th}$ and $25_{th}$ percentiles respectively for the results of 100-cross validation;
  The red straight line inside the box denotes the median value, while the blue dot denotes the mean value;
  The red pluses denote the outlier values.
  }
  \label{Bonn_all}
\end{figure*}
 
\begin{table}[h]
\caption{The mean accuracy of seven classification cases with four methods evaluated on \rev{Bonn dataset}.}
\begin{adjustbox}{width=3.6 in,center}
\centering
\begin{tabular}{cccccccccc}
\hline
\hline	
 Methods & Case 1	 & Case 2	 & Case 3 	& Case 4 	 & Case 5 	 & Case 6	 & Case 7	\\
\hline
MS-WTC	 & \textbf{1.000 }	 & \textbf{1.000 }	 & \textbf{0.998 }	 & \textbf{0.998 }	 & \textbf{0.998 }	 & \textbf{0.998} 	 & \textbf{0.999 }	\\
 M-WTC	 & 1.000 	 & 1.000 	 & 0.995 	 & 0.993 	 & 0.997 	 & 0.997 	 & 0.998 	\\
 S-WTC	 & 1.000 	 & 1.000 	 & 0.992 	 & 0.988 	 & 0.996 	 & 0.995 	 & 0.998 	\\
 FFT	 & 0.995 	 & 0.999 	 & 0.989 	 & 0.988 	 & 0.994 	 & 0.992 	 & 0.993 	\\
\hline	
\hline
\end{tabular}
\label{taba}
\end{adjustbox}
\end{table}

Our proposed MS-WTC method using the significant low-dimensional spectral-temporal features is examined via standard statistical measurements--SPE, SEN and ACC on \rev{Bonn dataset}, with the evaluation results compared with other methods using spectral features presented in Fig.~\ref{Bonn_all} and Tab.~\ref{taba}.
Examining results for the proposed method are cyan shadowed in Fig.~\ref{Bonn_all} and presented as bold in Tab.~\ref{taba}. 
All measurements on the MS-WTC method reach to nearly 100\%, except for a few outliers. 
Generally, using our MS-WTC method, the mean values of SPE, SEN and ACC are almost 100\% for seven cases, as shown by the blue dots in Fig.~\ref{Bonn_all}.

To demonstrate the better performance of our MS-WTC method, we also examine other methods using similar features on \rev{Bonn dataset}, such as M-WTC, S-WTC and FFT, with the results shown in Fig.~\ref{Bonn_all} and Tab.~\ref{taba}.
It is apparent from the results that the MS-WTC method has the best performance.
Actually, SPE of all methods here can stably reach to 100\%, while other methods are not so stable as our MS-WTC method on the performance of SEN as well as the resulting ACC (Fig.~\ref{Bonn_all}).   
In addition, the M-WTC and S-WTC method perform better than FFT with higher and more stable SEN as well as ACC.

Table~\ref{tabc} presents a comparison between the proposed classification scheme and the state-of-the-art methods published for seizure detection on \rev{Bonn dataset}.
The same testing dataset ensures that the comparison is persuasive and feasible. 
It can be observed from Tab.~\ref{tabc} that the proposed classification approach presents comparable classification performance among other existing methods. 
For example, the wavelet-based features are applied to neural network and SVM, which can produce accuracy of $100\%$ and $99.54\%$ in Case 7, respectively~\cite{xie2013wavelet,d64}. In comparison, the accuracy of $99.9\%$ can be achieved by using our proposed method in case 7.
The line length feature from DWT applied into ANN and the temporal features from entropy applied to the SVM can reach an accuracy of $97.75\%$ and $98.23\%$ in case 5 and 6, respectively~\cite{d61}, while the accuracy ($99.8\%$) for the proposed method in both case is slightly higher than existing works.
All other methods based on EEG signal pattern, such as entropy, covariance matrix and dual-tree complex wavelet transform, can achieve good performance on seizure detection, where our proposed method can also reach almost $100\%$ accuracy.
Overall, the proposed MS-WTC method achieves high mean accuracy (Tab.~\ref{tabc}) and is more stable (Fig.~\ref{Bonn_all}) than existing methods on \rev{Bonn dataset}. 

\begin{table*}[htbp]
\caption{Comparison of various seizure detection methods on \rev{Bonn dataset}}
\begin{adjustbox}{width=6 in,center}

\centering
\begin{tabular}{ccccccccccc}
\hline
\hline
Cases           &Year               & Method                                                & Accuracy (\%)     & Reference\\
\hline
Case 1          &2016               &EMD based temporal and spectral features and ANN       &99               &Riaz et al. \cite{riaz2015emd}\\
                &2014               &Fuzzy ApEn and linear SVM                              &100             & Kumar et al. \cite{d50}\\
                &\rev{2014}               &\rev{STFT and multilayer perceptron}                         &\rev{99.8}             & \rev{Samiee et al. \cite{samiee2014epileptic}}\\
                &\textbf{——}        &\textbf{MS-WTC}                                        &\textbf{100.0}      &\textbf{Proposed work }\\
\hline	
Case 2         &\rev{2014}               &\rev{STFT and multilayer perceptron}                         &\rev{99.3}             & \rev{Samiee et al. \cite{samiee2014epileptic}}\\

                &2018               &MRBF-MPSO-OLS algorithm and SVM           	            &99.3	            &Li et al. \cite{d56}\\
                &2020               &Covariance matrix and AB-LS-SVM		                &100                &AI-Hadeethi et al. \cite{d53}\\
                &\textbf{——}        &\textbf{MS-WTC}                                        &\textbf{100.0}     &\textbf{Proposed work }\\
\hline
Case 3          &2014               &Horizontal visible graph mapping of EEG signals and KNN&99.6               &	Zhu et al. \cite{d57}\\
                &2016               &Dual-tree complex wavelet transform                    &100                &	Swami et al. \cite{d67} \\
                &                   &and a general regression neural network                &                   &\\
                &2018               &Complete ensemble EMD and Adaboost                     &	99              &	Hassan et al. \cite{d58}\\
                &\textbf{——}        &\textbf{MS-WTC}                                        &	\textbf{99.8}   &	\textbf{Proposed work }\\
\hline	
Case 4          &2020               &Segmented EEG signals with novel CNN                   &	97.63           &	 Zhao et al. \cite{d52}\\
                &2020               &Covariance matrix and AB-LS-SVM                        &	99              &	AI-Hadeethi et al. \cite{d53}\\
                &\textbf{——}        &\textbf{MS-WTC}                                        &	\textbf{99.8}   &	\textbf{Proposed work }\\
\hline	
Case 5          &2010               &Line length feature from DWT and ANN                   &	97.75           &	Guo et al. \cite{d61}\\
                &2014               &Fuzzy ApEn and linear SVM                              &	98.15           &	Kumar et al. \cite{d50}  \\
                &\textbf{——}        &\textbf{MS-WTC}                                        &	\textbf{99.8}   &	\textbf{Proposed work }\\
\hline	
Case 6          &2014               &Fuzzy ApEn and linear SVM                              &	98.23           &	Kumar et al. \cite{d50} \\
                &\textbf{——}        &\textbf{MS-WTC}                                        &	\textbf{99.8}   &	\textbf{Proposed work }\\
\hline
Case 7          &2013               &Genetic programming                                    &	100            &	Fern{\'a}ndez-Blanco et al. \cite{fernandez2013classification}\\
                &2013               &Wavelet-based sparse function and neural network	    &   100          &	Xie et al. \cite{xie2013wavelet}\\
                &2016               &DWT followed by PCA and SVM                            &	99.54           &	Wang et al. \cite{d64}\\
                &2021              &Fuzzification with fuzzy decision tree and fuzzy random forest	    &   99.5            &	Jan et al. \cite{9444205}\\
                &\textbf{——}        &\textbf{MS-WTC }                                       &	\textbf{99.9}   &	\textbf{Proposed work }\\
\hline
\hline
\end{tabular}
\label{tabc}
\end{adjustbox}
\end{table*}

\subsection{Results for \rev{C301 dataset}}
\label{Results for database B}

\begin{figure}[h]
  \centering
  \includegraphics[width=1\columnwidth,trim={0 0 0 0}, clip]{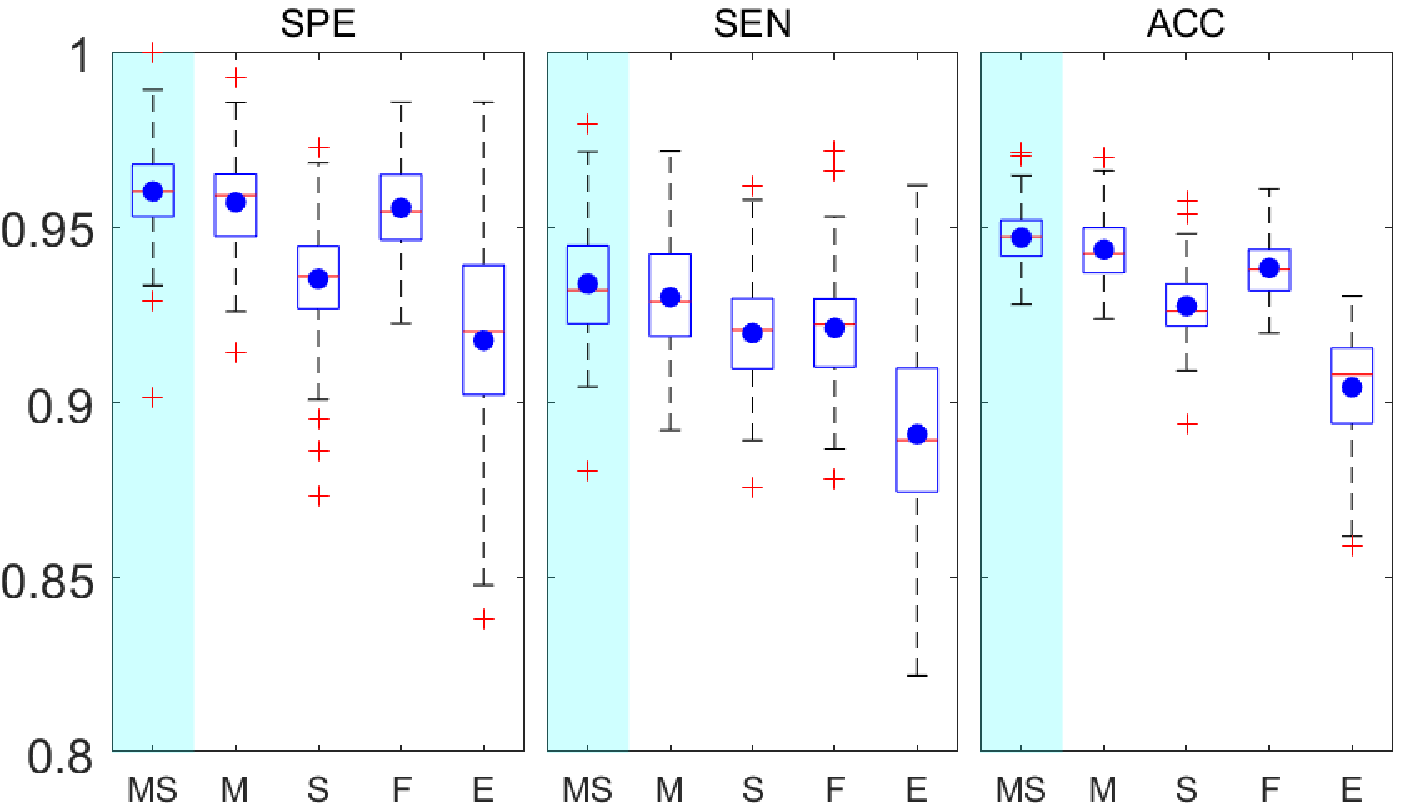}
\caption{Evaluations of various methods: MS-WTC (MS), M-WTC (M), S-WTC (S), FFT (F) and EMD (E) with 100-cross validation on \rev{C301 dataset for mixed scenario.}}
 \label{301_all_mix}
\end{figure}

\begin{figure}[h]
  \centering
  \includegraphics[width=1\columnwidth,trim={0 0 0 0}, clip]{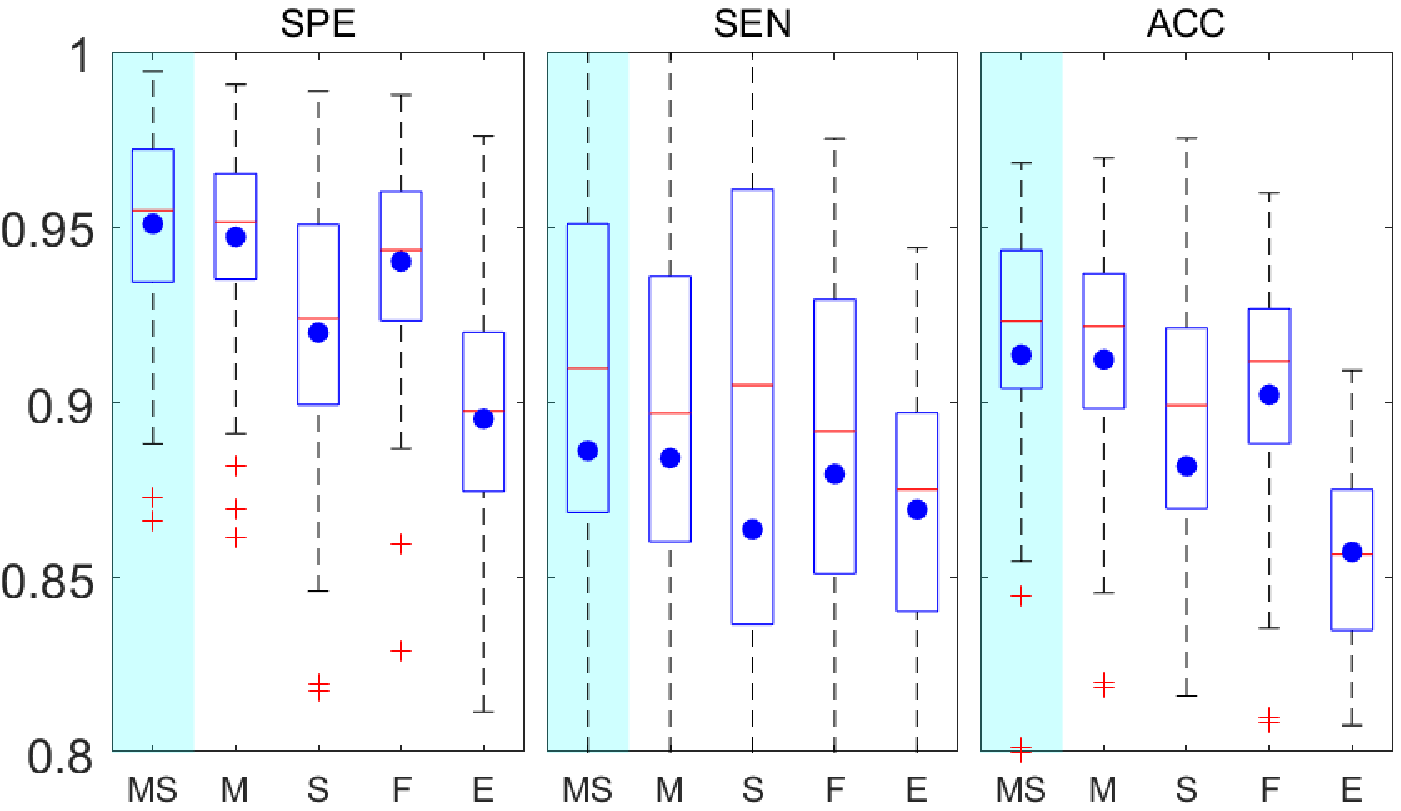}
\caption{Evaluations of various methods: MS-WTC (MS), M-WTC (M), S-WTC (S), FFT (F) and EMD (E) with 100-cross validation on \rev{C301 dataset for separated scenario.}}
 \label{301_all_sep}
\end{figure}

\begin{table}[h]
\caption{Comparison of various seizure detection methods on \rev{C301 dataset for mixed scenario.}}
\begin{adjustbox}{width=3.2 in,center}
 
\centering
\begin{tabular}{cccccc}
\hline
\hline	
    Methods	             & Measurements		& Mean	            	         & Standard deviation\\
\hline							
	\textbf{MS-WTC}	     & \textbf{SPE}	    & \textbf{0.960} 	 	         & \textbf{1.39E-02}\\
		                 & \textbf{SEN}	    & \textbf{0.934} 	 	         & \textbf{1.60E-02}\\
		                 & \textbf{ACC}	    & \textbf{0.947} 	  	         & \textbf{8.33E-03}\\
\hline				
	M-WTC	             & SPE	            & 0.957 	         	         & 1.43E-02	\\
		                 & SEN	            & 0.930 	          	         & 1.55E-02	\\
		                 & ACC	            & 0.944 	          	         & 8.99E-03	\\
\hline				
	S-WTC	             & SPE	            & 0.935 	          	         & 1.64E-02	\\
		                 & SEN	            & 0.920 	        	         & 1.50E-02	\\
		                 & ACC	            & 0.928 	        	         & 9.6E-03	\\
\hline				
	FFT	                 & SPE	            & 0.955 	         	         & 1.33E-02	\\
		                 & SEN	            & 0.921 	         	         & 1.57E-02	\\
		                 & ACC	            & 0.938 	         	         & 8.67E-03	\\
\hline				
	EMD	                 & SPE	            & 0.918 	         	         & 2.98E-02	\\
		                 & SEN	            & 0.891 	          	         & 2.75E-02	\\
		                 & ACC	            & 0.904 	         	         & 1.53E-02	\\
\hline
\hline	
\end{tabular}
\label{tab_mix}
\end{adjustbox}
\end{table}

\begin{table}[h]
\caption{Comparison of various seizure detection methods on \rev{C301 dataset for separated scenario.}}
\begin{adjustbox}{width=3.2 in,center}
 
\centering
\begin{tabular}{cccccc}
\hline
\hline	
    Methods	             & Measurements		& Mean	            	         & Standard deviation\\
\hline							
	\textbf{MS-WTC}	     & \textbf{SPE}	    & \textbf{0.951} 	 	         & \textbf{2.74E-02}\\
		                 & \textbf{SEN}	    & \textbf{0.886} 	 	         & \textbf{1.06E-01}\\
		                 & \textbf{ACC}	    & \textbf{0.914} 	  	         & \textbf{4.87E-02}\\
\hline				
	M-WTC	             & SPE	            & 0.947 	         	         & 2.72E-02	\\
		                 & SEN	            & 0.884 	          	         & 9.64E-02	\\
		                 & ACC	            & 0.912 	          	         & 4.52E-02	\\
\hline				
	S-WTC	             & SPE	            & 0.920 	          	         & 4.04E-02	\\
		                 & SEN	            & 0.864 	        	         & 1.43E-01	\\
		                 & ACC	            & 0.882 	        	         & 6.39E-02	\\
\hline				
	FFT	                 & SPE	            & 0.940 	         	         & 2.78E-02	\\
		                 & SEN	            & 0.880 	         	         & 8.30E-02	\\
		                 & ACC	            & 0.902 	         	         & 4.52E-02	\\
\hline				
	EMD	                 & SPE	            & 0.895 	         	         & 3.45E-02	\\
		                 & SEN	            & 0.896	          	             & 3.84E-02	\\
		                 & ACC	            & 0.857 	         	         & 2.55E-02	\\
\hline
\hline	
\end{tabular}
\label{tab_sep}
\end{adjustbox}
\end{table}

\black{
To demonstrate the reliability and stability of the proposed MS-WTC method, we further examine it on \rev{C301 dataset} in terms of SPE, SEN and ACC, compared with other methods: M-WTC, S-WTC, FFT and EMD. \rev{Additionally, as mentioned in section \ref{Training and classification}, we tested mixed scenario and separated scenario to complete research results.}} 
The results are shown in Fig.~\ref{301_all_mix}\rev{, Fig.~\ref{301_all_sep},} Tab.~\ref{tab_mix} and \rev{Tab.~\ref{tab_sep}}, which quantifies the mean and standard deviation of these measurements. 

Figure~\ref{301_all_mix} shows that our MS-WTC method outperforms others in terms of the mean SPE, SEN and ACC.
The mean ACC of our method can reach to $94.7\%$ (Tab.~\ref{tab_mix}), which is significantly better than that of EMD based on the temporal features ($90.4\%$).
The simple spectral feature by FFT can reach a comparable SPE ($95.5\%$) with our MS-WTC method ($96.0\%$), but its SEN ($92.1\%$) is relatively lower than ours ($93.4\%$), resulting in a relatively lower ACC ($93.8\%$).
Noticeably, the performance of the M-WTC method using only the time-averaged PSD magnitude is comparable with the MS-WTC method on this dataset, which is out of reach for the S-WTC method (Fig.~\ref{301_all_mix}).
Besides, our MS-WTC method is the most stable one for seizure detection, comparing to all other methods, as indicated by the smallest box size in Fig.~\ref{301_all_mix} and smallest standard deviation ($8.33E-03$) of ACC presented in Tab.~\ref{tab_mix}.

\rev{Figure~\ref{301_all_sep} also shows that our MS-WTC method outperforms others in terms of the mean SPE, SEN and ACC. The mean ACC of our} \rev{method can reach to $91.4\%$ (Tab.~\ref{tab_sep})}. 

\rev{The results of both scenarios are presented, and as expected, the mixed scenario indeed performs better with more stable results (shorter boxed in Fig.~\ref{301_all_mix} and smaller Standard deviation values in Tab.~\ref{tab_mix}) due to the smaller differences among cross-validation sets. However, the results of the separated scenario are comparable in terms of mean accuracy (ACC=0.914 for the MS-WTC method in Tab.~\ref{tab_sep}), which is slightly lower than that (ACC=0.947) of the mixed scenario. Thus, our proposed MS-WTC method can achieve competitive results for both scenarios, and then clinically applicable.} 

\black{The performance results demonstrate that our MS-WTC method is accurate in seizure detection on the published \rev{Bonn dataset} as well as the clinical \rev{C301 dataset}.
These validate that our proposed low-dimensional spectral-temporal features are significant for seizure detection.

\section{\label{Conclusion}Conclusion}

In this study, we discovered the significant low-dimensional spectral-temporal features for seizure onset detection.
These features capture the intrinsic rhythmic and non-stationary properties of seizure EEG signals, and simultaneously reduce redundant information.
Using these features, we developed a CWT-based compact MS-WTC method, which is validated as a reliable, efficient and stable method in seizure detection, by comparison with other methods using low-dimensional temporal and spectral features, such as EMD and FFT.
The results of our method are verified on two datasets: the widely-used benchmark Bonn dataset~\cite{d23} (\rev{Bonn dataset}) and a clinical EEG dataset from the Chinese 301 Hospital (\rev{C301 dataset}), which is interfered with the physiological and non-physiological artifacts.
For  the former,  seven  classification  tasks  were  evaluated  with  the  accuracy  from  $99.8\%$  to  $100.0\%$,  while  for  the  latter,  the  MS-WTC method achieved the average accuracy at $94.7\%$ \rev{for mixed scenario}, outperforming other methods with low-dimensional temporal and spectral features. 
This indicates the robustness of our method, and confirms the significance of the extracted low-dimensional spectral-temporal features.

These significant features based on CWT consist of the time-averaged PSD magnitude ($\mu$) and its standard deviation ($\sigma$) at each time scale, which are integrated and fed into a CNN model for classification. 
Based on the experimental results in Fig.~\ref{Bonn_all}, Fig.~\ref{301_all_mix} \rev{and Fig.~\ref{301_all_sep}}, our MS-WTC method using these two features outperforms the M-WTC and S-WTC methods using only one of them, verified on both datasets.
It seems that the M-WTC method is more significant than the S-WTC method in seizure detection on \rev{C301 dataset}, while they achieved a considerable performance on \rev{Bonn dataset}.
Overall, their combination is effective in seizure detection.

To this end, we have successfully extracted the significant and compact statistical information inside the time-frequency image, which has been shown to be sufficient and powerful feature information for seizure detection as well as prediction~\cite{d12,d10,d75}.
Our proposed features not only capture the intrinsic properties of non-stationary seizure EEG signals, but also effectively reduce the feature dimension of the 2D image produced by CWT. \rev{The high-dimensional spectral-temporal features by CWT is not examined here due to large computational requirements and its cone-shape image with
boundary effects excluded. Moreover, for the more complicated and much larger dataset, such as the CHB-MIT dataset \cite{d66} collected at Children’s Hospital Boston database, the 2D image feature by CWT requires heavy computational load, while our method proposed here
can facilitate such kind of big data analysis by reducing redundant information and capturing the essentials. These two issues will be further explored in our future work.} 

In conclusion, the significant low-dimensional spectral-temporal features were effectively extracted for seizure detection, both on the published dataset as well as the clinical raw EEG dataset.
Our proposed MS-WTC method is reliable, efficient and stable in seizure detection, with the potential for clinical use.
\bibliographystyle{IEEEtran}
\bibliography{reference_EEG}}

\end{document}